\documentstyle[aps,pre,multicol,epsfig]{revtex}

\input{epsf}

\begin{document}

\date{21 december 1998}
\title{Enhancement of Andreev reflection by spin polarization
in \\ out-of-equilibrium ferromagnet-superconductor junctions}
\author{O. Bourgeois$^{1}$, P. Gandit$^{1}$, J. Lesueur$^{2}$, R.
M\'elin$^{1}$, A. Sulpice$^{1}$, X. Grison$^{2}$ and J. Chaussy$^{1}$}
\address{$^{1}$Centre de Recherches sur les Tr\`es Basses
Temp\'eratures (CRTBT), CNRS, associated to the Universit\'e Joseph Fourier,\\
BP166, 38042
Grenoble CEDEX 9, France \\$^{2}$Centre de Spectrom\'etrie Nucleaire et de
Spectrom\'etrie de Masse CNRS-IN2P3,\\ Universit\'e Paris Sud, Bat 108, 91405
Orsay, France }
\maketitle
\begin{abstract}
We report on transport measurements on Nb/Al/Gd/Al/Nb junctions.
Bulk Gadolinium is a weakly polarized ferromagnet (5-7\%), and is
present in the junction in granular form (superparamagnet).
We show that Andreev reflection is strongly
enhanced by a weak polarization, obtained
by applying an external magnetic field.
A new model is proposed that accounts for this effect.
The transport is described in terms of ``hot'' carriers
that experience Zeeman splitting due to the spin polarized
background.
The Landauer formula is $G={{4e^2} \over
h}A(h_{\rm b})$, $h_{\rm b}$ being the exchange field
at the grain boundary,
and A(E)
the Andreev current coefficient, that increases with E for resistive
junctions.
\end{abstract}

\centerline{PACS numbers:74.50.+r, 74.80.Dm, 75.70.Cn, 85.30.St}

\bigskip
\begin{multicols}{2}
\narrowtext
Much interest has been recently devoted to understand the interplay
between magnetism and superconductivity. For example, recent experiments on
ferromagnet-superconductor superlattices have shown anomalous oscillating
critical temperatures \cite{jiang95,2muhg98} due to a possible coupling
between the magnetic layers through the superconductor. Other
experiments have shown new coherence phenomena in
superconducting-ferromagnetic
systems \cite{petra94,lo96,monic} directly related to the presence of
magnetic ordering.
The problem which arises is to clearly understand the behavior of
normal metal-superconductor (N/S) junctions in the presence of
ferromagnetism and more specifically, the importance of
spin polarization on Andreev reflection in such junctions.
In a N/S sandwich,
the zero temperature current is transmitted below the
superconducting gap via Andreev reflection only : an
incoming electron with a spin $\sigma$
is reflected at the interface as a hole in the band of
opposite spin $- \sigma$, whereas a Cooper pair
is transferred into the superconductor \cite{andreev64}.
Blonder, Tinkham and Klapwijk (BTK) \cite{btk}
have shown that Andreev reflection is weakened
when interfacial elastic processes
are taken into account under the form of a repulsive
potential $V(x) = H \delta(x)$ at the interface. This theory
interpolates between a perfect transparent interface and an insulating
barrier as the dimensionless barrier strength $Z=H/ \hbar
v_F$ increases from zero to infinity.
It is of importance to understand the behavior of such junctions
when the normal metal is ferromagnetic.
In this case, assuming thermal equilibrium within the ferromagnet, one
spin population is depleted with respect to the other.
Considering $N_{\uparrow}$ channels in the spin up
band, and $N_{\downarrow}<N_{\uparrow}$ channels in the spin down band,
de Jong and Beenakker \cite{jongbee95} noticed
that only $N_{\downarrow}$ channels are
available for Andreev reflection . The corresponding Landauer
formula is $G=4 N_{\downarrow} e^2/h$.
Andreev reflection is
thus strongly dependent on the Fermi surface spin polarization.

Recent experiments in ferromagnet-superconductor (FM/S)
junctions by Soulen {\it et~al.}~\cite{soulen98}
(with a ferromagnetic metal
FM=Ni$_{0.8}$Fe$_{0.2}$, Ni, Co, Fe, NiMnSb,
La$_{0.7}$Sr$_{0.3}$MnO$_3$) and by Upadhyay {\it et~al.}\cite{up98}
(with FM=Co, Ni), have proved that Andreev reflection is strongly
suppressed as the Fermi surface polarization is increased,
in full agreement with the predictions by de Jong and
Beenakker \cite{jongbee95}. For instance, 
the low temperature and 
low voltage differential conductance $G_{S}(0)=(d G(V)/dV)/G_n$ 
in N/S junction
is twice its value above the superconducting gap when N is a non 
magnetic material, because of Andreev reflection. On the other 
hand, $G_{S}(0)$ falls to zero in the most polarized FM/S junctions 
(90~\% polarized if FM=CrO$_2$), showing the suppression
of Andreev reflection by spin polarization~\cite{soulen98,up98}.

The aim of the present Letter is to show that spin polarization
can enhance Andreev reflection in S/FM/S junctions with a
sizeable interfacial scattering
($Z\approx 2$). This provides the opportunity of experimenting FM/S
junctions in between a point contact regime with $Z\approx 0$
\cite{soulen98,up98} and a tunneling regime with $Z\gg 1$ \cite{tm}.
We used Gadolinium
(Gd) as ferromagnetic metal, in which the conduction band
is weakly polarized, between
5 and 7~\%  \cite{mpt,busch}; a small value
compared to the polarization of 37~\% in {\it e.g.}
NiFe reported in the less polarized sample studied in
Ref. \cite{soulen98}.
Unlike the transition metals with s-d itinerant magnetism, 
Gd has 4f localized moments, ferromagnetically aligned through the
RKKY interaction mediated
by the conduction electrons. In spite of the complexity of the
conduction band structure, we believe that the physics involved in
our experiment can be captured by assuming a Stoner model for the
conduction band of width $W\approx$10~eV \cite{suhl} in the presence
of an exchange field $h_{\rm Gd}$
of the order of a 140~meV for bulk Gd \cite{coq}.

A magnetic field applied parallel to the junction
is used to tune the spin polarization.
We show that even with polarizations as small as $5-7 \%$ in
the conduction band of bulk Gd, the low voltage resistance
of the junction is strongly dependent on its
polarization. More specifically, the resistance of the junction
{\it decreases} when an external magnetic field
is applied. We demonstrate that this behavior
originates from an enhancement of Andreev
reflection when the boundary spin polarization of the granular
Gd increases.
We set up a new model which accounts for
our observations. It is based on the
transport of ``hot'' electrons in the Gd grains, which do not
thermalize with the background ferromagnet conduction band,
but experience Zeeman splitting.
This model relies on a reduction of the Gd magnetism
at the grain boundaries.

\begin{figure}[tb]
        \epsfxsize=\columnwidth
        \centerline{\epsfbox{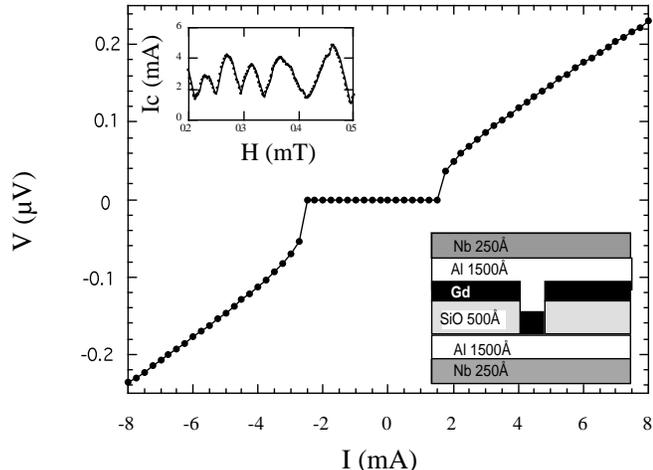}}
        \caption{Current-voltage characteristics at T=2 K
with a thickness of $40$~\AA \ Gd. The top left inset shows
the modulation of the critical current under low magnetic field. 
A schematic sample
cross section is shown on the bottom right inset. All samples with
different Gd thicknesses exhibit a similar I-V characteristic. The
magnetic field is applied in the plane of the junction.}
        \label{V(I)}
        \end{figure}

We study the specific
junction Nb/Al/Gd/Al/Nb. The layers are prepared by 
e-beam evaporation in ultra-high-vacuum 
on a silicon substrate held at room
temperature. Layer thicknesses are controlled
during growth with a quartz crystal monitor.
A typical rate of 1~\AA /s is used for
Gd. The base pressure is below
$3\times10^{-9}$ Torr,
and the working pressure is $2\times10^{-9}$ Torr for Gd,
$4\times10^{-9}$ Torr
for Al and Nb.
The junction area ($0.1\times 0.1$~mm$^{2}$) is defined 
by evaporating insulating SiO through shadow masks at $2\times10^{-8}$ 
Torr typically. The inset of Fig.~\ref{V(I)} shows a schematic
cross section of the junction, which is made of two Nb leads
of 250~\AA,
and a sandwich Al/Gd/Al with a Gd thickness ranging from 20 and 100~\AA.
A 1500~\AA \ Al layer has been used to locally lower the 
superconducting $T_{c}$ of the junction and avoid spurious effects 
due to the Nb leads transitions.

\begin{figure}[tb]
        \epsfxsize=\columnwidth
        \centerline{\epsfbox{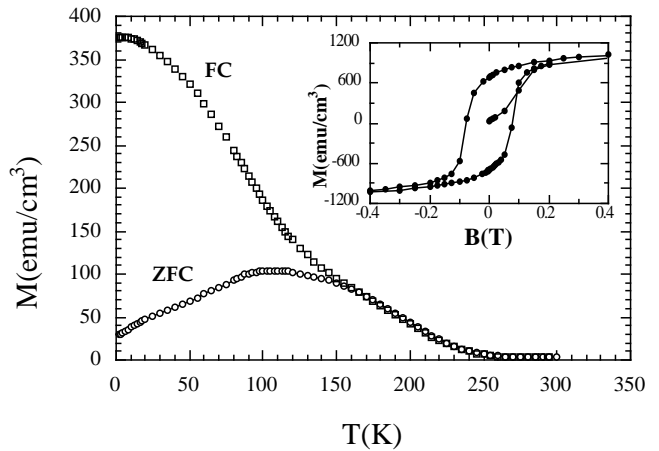}}
        \caption{ Magnetization of
a 100~\AA \ Gd layer versus temperature
measured in a ZFC-FC procedure in low magnetic field H=50~Oe.
The Nb layer is not present in this measurement.
Inset : hysteresis loop of the same sample at 2~K.}
        \label{M(T)}
        \end{figure}

Several methods were used to characterize the Gd layer. First,
Transmission Electron Microscopy images on Al/Gd reference samples
evaporated under the same conditions as the junctions, have shown
that Gd layers thinner than 20~\AA \ are not continuous, thus imposing 
a lower bound on samples thicknesses. Gd
is polycrystalline and the size of grains
ranges between 70 to 100~\AA. For the smaller thicknesses,
the grains are elliptic, of typical dimension
20 \AA $\times$ 70 \AA $\times$ 70 \AA.
The magnetic properties of Gd
were investigated by means of SQUID magnetometry. In Fig.~\ref{M(T)},
we show a Zero Field Cooled (ZFC) and Field Cooled (FC)
 magnetization curve in a low applied field (H=50~Oe).
These curves exhibit a large thermomagnetic
hysteresis with an irreversibility temperature ($T_{i}$) of 150~K, and
a blocking temperature of about 120~K.
We estimate that the Curie temperature range between 50~K and 100~K.
This behavior is characteristic of
a superparamagnet, in agreement with the granularity
of the magnetic layer. From the
saturation magnetization, the magnetic moment per atom is estimated to
be $4\mu_{B}$,
substantially below the Gd bulk moment of $7.6\mu_{B}$.
We attribute this reduction
of the saturation magnetization compared to the bulk value
to frozen spins that do not align in the applied magnetic
field due to spin glass effects as it is expected for Gd small grains 
\cite{douglass,riegel}.
The coercitive
field of the Gd layers is around 100~mT, for all
the Gd thickness.
The macroscopic magnetization of the granular Gd
increases as the external field is increased
because the giant spin grains align on average
in the external field.

\begin{figure}[tb]
        \epsfxsize=\columnwidth
        \centerline{\epsfbox{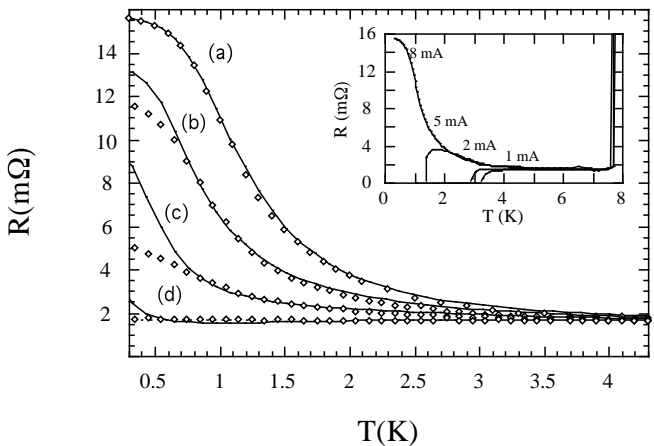}}
        \caption{ Ohmic resistance versus temperature
measured with a bias
current larger than twice the critical current under a magnetic
field applied parallel to the layers (a) H=0~mT, (b) 40~mT, (c) 60~mT, (d) 100~mT. 
The solid lines show the fits
to the N/FM/S model. We have used $Z=2.3$  and
$h_{\rm b}=0$, $0.38$, $0.61$, and $0.83$ respectively,
in units of the Al superconducting gap of the order of a meV.
The inset shows the behavior of the transition
temperature under different bias currents in zero external
field.}
        \label{R(T)}
        \end{figure}

All the transport measurements have been performed with a highly sensitive
dc-current
method using a four probe technique \cite{dauguet}. Resistances range 
from a few m$\Omega$ to a few $\mu \Omega$ and the voltage across the junction
 is of the order of a nanovolt. A typical current-voltage I-V
characteristics is shown on Fig.~\ref{V(I)}. It displays a sizeable 
critical current I$_c$, whose amplitude strongly decreases with the Gd 
thickness, indicating a Josephson coupling between the superconducting 
electrodes through the ferromagnet. This will be 
exposed in details in a forthcoming publication. The modulation of 
I$_c$ in a weak applied magnetic field (inset of Fig.~\ref{V(I)}) 
confirms the existence of a Josephson coupling and excludes
that pinholes dominate the 
transport properties.
In zero magnetic field (see Fig.~\ref{R(T)}) 
a first transition appears in transport
measurements corresponding to the superconducting
transition of Nb between 7.5 and 8~K, that induces superconductivity
in Al by proximity. At lower temperatures the resistivity vanishes due to
Josephson effect (inset of Fig.~\ref{R(T)}).
Fig.~\ref{R(T)} shows the
effect of a magnetic field on the junction resistance, the magnetic
field being applied along the layer and perpendicular to the current.
The resistance is drastically {\it reduced} by the external magnetic
field.
We can exclude that this behavior originates from part of the bilayer
Nb/Al becoming normal. The parallel critical field ($H_{//}$)
in a thin film is
$H_{// }=\sqrt {24}{{\lambda (t)} \over d}H_c$, where $\lambda (t)$
is the penetration depth in the superconductor and $d$ the film
thickness\cite{tin}. For Niobium
this leads to $H_{//\rm Nb}=1.7$~T, a value much larger than the 
fields used in this investigation. As for Al, the magnetic field required
to destroy superconducting proximity effect is larger than the bulk
critical field $H_{c2}\approx $ 200~mT of Nb \cite{hur66}. Therefore,
the bilayer remains superconducting for a magnetic field up to 100~mT.

\begin{figure}[tb]
        \epsfxsize=\columnwidth
        \centerline{\epsfbox{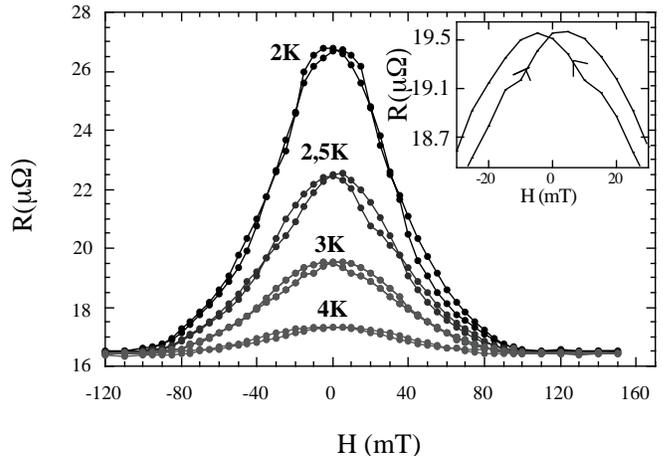}}
        \caption{Negative magnetoresistance obtained
for different temperatures on a sample with a
40~\AA \ Gd layer and a bias current of 5~mA.
The details of the
hysteresis at 3~K are shown in the inset, the arrows indicating the
direction in which the field is swept.}
        \label{R(H)}
        \end{figure}

Fig.~\ref{R(H)} shows the variations of the ohmic resistance at a
fixed temperature as a function of the applied magnetic field. The
sample is first zero field cooled from 10~K to the measurements 
temperatures. The magnetic field is swept from
-150~mT up to 150~mT, and back to -150~mT. The crossover field for the
negative magnetoresistance is independent on temperature, which is an
additional proof that this effect does not originate from the 
destruction of
superconductivity in the Nb/Al bilayer, in which case this crossover
field would depend on temperature. Moreover, the hysteresis in the
magnetoresistance clearly indicates the onset of a magnetic effect
in the transport in the
junction (inset of Fig.~\ref{R(H)}).

The de Jong-Beenakker theory cannot account for the present
observations, since it would predict an increase of
resistance upon spin polarization, and also because the relative
change in resistance induced by a small spin polarization of 5-7~\%
would be of the order of the spin polarization \cite{theor},
incompatible with the magnitude of the resistance variations shown on
Fig.~\ref{R(T)}.
We believe
that the failure of the de Jong-Beenakker theory in
our system lies in the fact that transport in the Gd
grains is dominantly an out-of-equilibrium transport
since the grains (70-100~\AA) and the Gd thicknesses are small.
We present here a modeling of this situation that
qualitatively accounts for the negative magnetoresistance.
In this model, the charge carriers are ``hot'' electrons,
which wave vector and spin remain constant while they
cross the grains. Given the small size of the grain, we neglect any 
inelastic scattering. The spin
energy of an electron (hole) is $-(+)h_{\rm Gd}({\bf x})
\sigma$, $h_{\rm Gd}({\bf x})$ being the space-dependent
exchange field. The relevant quantity for Andreev reflection
is the exchange field $h_{\rm b}$ at the grain boundary,
a quantity which can be much smaller than the bulk
value $h_{\rm Gd} \simeq 140$~meV. The magnetism of Gd is very 
sensitive to growth parameters
\cite{farle93,lohn98}, and surface magnetization can be strongly 
weakened by 
possible adsorbtion of oxygen atoms at the edge of Gd grains 
\cite{macilroy96A}.
An applied magnetic field will favor ferromagnetism
at the boundaries and increase $h_{\rm b}$. As the
external field increases, the nucleation of boundary
polarization is expected to favor the reversal of the bulk
giant moment if the grain had a misoriented
magnetization. This Gd grain modeling
also allows to understand that the hysteresis
in resistance is weaker than in magnetization,
since the former involves boundary
magnetism and the latter the superparamagnetic
behavior of the giant spin grains.

Now, let us consider the semiclassical treatment of a
N/FM/S junction  \cite{otbk}, and assume a scattering
only at the FM/S interface I$_2$.
The presence of a scattering at both interfaces
leads to similar qualitative conclusions, as well
as the S/FM/S junction, in which cases
a numerical treatment is necessary and will
be presented elsewhere\cite{theor}.
We note
$f_{R,(L)}^{(\sigma )}(E,I_{1,(2)})$ the non-equilibrium
distribution functions of right (left)
movers with a spin $\sigma$ at
the I$_1$=N/FM
(I$_2$=FM/S) interface. The Zeeman splitting in the
FM region is taken into account by proper boundary
conditions, {\it e. g.}
$f_{R}^{(\sigma )}(E,I_1)=f_T(E+ h_{\rm b} \sigma)$
at the I$_1$ interface, $f_T$ being the equilibrium distribution function 
in the reservoirs. The
exchange field and the superconducting gap
are assumed to have step function variations at the
interface I$_2$. We have point split the two
order parameters at this interface by
assuming that electrons incoming from
the S part will first experience the barrier
potiential and next Zeeman splitting, or the
opposite. In the first case, the barrier
conductance is independent on spin polarization,
whereas in the second situation, in which both
orders coexist close to the interface, the
I-V relation is
\begin{eqnarray}
I & = & {e \over h}\int \left( 1+A(E)-B(E) \right)
\left[ f_T(E-eV-h_{\rm b}) \right. \\
&&\left. +f_T(E-eV+h_{\rm b})-f_T(E-h_{\rm b}) -
f_T(E+h_{\rm b})
\right] \nonumber,
\end{eqnarray}
 with $A(E)$ and $B(E)$ are the Andreev and backscattered transmission 
 coefficients \cite{btk}.
The Landauer formula is
$$G=\left({4e^2}/ h \right)A(h_{\rm b}),$$
a quantity that increases with the spin
polarization $h_{\rm b}$ if $Z\approx 2$ \cite{btk}. 
As shown on Fig.~\ref{R(T)}, the best fits
are in a qualitative agreement with experiments.
Because of the $\sim 5$~mT hysteresis in
conductance, the spontaneous boundary magnetization is expected
to be finite. However
we cannot resolve magnetic fields
below $\sim 5$~mT whithin the present model.

Given the weakening of ferromagnetism at the grain
boundaries \cite{farle93,lohn98}, we have shown that
the observed negative
magnetoresistance originates from an enhancement of Andreev
reflection due to Zeemann splitting
of ``hot'' carriers.
This requires the coexistence of superconductivity
and ferromagnetism close to the interface.
A more complete understanding of this interface would
involve a treatment of the influence of the ferromagnet onto
the superconductor,
and, in view of the reduction of magnetism, the possibility of
cryptomagnetic \cite{anderson59}, or
cryptomagnetic-like structures \cite{buzdin88}
at the grain boundaries.

The authors thank M. Aprili, O. Buisson,
P. Butaud, C. Chappert, H. Courtois, B. Dou\c{c}ot, M. Giroud,
F. Hekking, T. Klapwijk,
C. Lacroix, B. Lussier, K. Matho, B. Pannetier and Ned S. Wingreen
for fruitful discussions and their encouragements, and M.O. Ruault for 
the TEM images.

\end{multicols}
\end{document}